\documentclass[twocolumn,superscriptaddress]{revtex4}
\usepackage{graphicx}
\usepackage{bm}
\usepackage{color}
\usepackage{amsmath}
\usepackage{amsfonts}
\usepackage{amssymb}
\usepackage{natbib}
\usepackage{latexsym}
\usepackage{mathrsfs}

\begin{document}

\title {A Quantum Mechanical Justification of Bohr's Atomic Model}

\author{Sergio A. Hojman}
\email{sergio.hojman@uai.cl}
\affiliation{Departamento de Ciencias, Facultad de Artes Liberales,
Universidad Adolfo Ib\'a\~nez, Santiago 7491169, Chile.}
\affiliation{Departamento de F\'{\i}sica, Facultad de Ciencias, Universidad de Chile,
Santiago 7800003, Chile.}
\affiliation{Centro de Recursos Educativos Avanzados,
CREA, Santiago 7500018, Chile.}

\begin{abstract}
Bohr's atomic model is based on the assumption that electrons on allowed quantized orbits do not radiate. Its main results include the values of the radii of circular quantized orbits and of the hydrogen atom energy levels. Quantum mechanical justifications of both his hypothesis and of the well known radii relation, which is used to compute the energy levels, are presented. 
\end{abstract}

\maketitle
 
\section{Introduction}

Students who are for the first time introduced to Bohr's atomic model as part of a historical account of the development of quantum mechanics, are probably skeptical about how one may safely assume that electrons moving in the allowed circular orbits around the atomic nucleus do not radiate. They may also wonder about why the model's prediction of those orbital radii when used to compute the hydrogen energy levels turn out to coincide with both the quantum Hamiltonian energy eigenvalues and with experimental results.\\

The purpose of this article is to use conventional quantum mechanics to contribute to clarifying these issues, hopefully helping to relieve students doubts.\\

Let me start by recalling that, in his groundbreaking article \cite{bohr}, Bohr starts the second paragraph of Part I by stating {\it{``Let us at first assume that there is no energy radiation.''}} for electrons moving on the allowed quantized orbits, in order to be able to construct a stable model for atoms.\\

Its main results are the determination of the Bohr radius for the ground state circular orbit and the relation of the radii of circular quantized orbits to it and the prediction of the hydrogen atom energy levels which, in turn, lead to the solution of the discrete spectra of atoms problem and to the correct values of the frequency of the hydrogen atom emission and absorption lines.\\ 

It has been known for a long time that the correct prediction of the hydrogen energy levels has, of course, full agreement with the Schrödinger equation hydrogen atom Hamiltonian energy eigenvalues.\\

Nevertheless, as far as I know, no simple cogent, either {\it{a priori}} or {\it{a posteriori}}, quantum mechanical justifications of both the no--radiation assumption and of the result predicting the orbits' radii exist.\\

This work presents quantum mechanical justifications for both Bohr's hypothesis and the radii result based on the application of the Schrödinger equation.\\

\section{The Madelung--Bohm treatment of the Schrödinger equation}

Let us start by considering the Schrödinger equations for an arbitrary external real time--independent potential $V=V(\vec{r})$
\begin{eqnarray}\label{sch1}
\left(-\frac{\hbar^2}{2m} \vec{\nabla}^2 + V(\vec{r}) - i \hbar \frac{\partial}{\partial t} \right)  \Psi(\vec{r},t)&=&0\ ,\ \ \ \  {\text {and}}\nonumber \\
\left(-\frac{\hbar^2}{2m} \vec{\nabla}^2 + V(\vec{r}) + i \hbar \frac{\partial}{\partial t} \right)  {\Psi}^{*}(\vec{r},t)&=&0\ . \end{eqnarray}
The polar version of the wave function $\Psi(\vec{r},t)$  may be written in terms of $A(\vec{r},t)$ and $S(\vec{r},t)$ that are real functions of space and of time as \cite{mad,bohm,holland,wyatt,ha201,ha20,ha21,hams21,ahms21},
\begin{equation}\label{psi}
{{\Psi}}(\vec{r},t) = A(\vec{r},t)\ e^{i S(\vec{r},t)/\hbar}\, .     \end{equation}
Now, write the real and imaginary parts of the Schrödinger equations respectively as,
\begin{eqnarray}
\frac{1}{2m} ({\vec{\nabla}S})^2 - \frac{\hbar^2}{2m}\frac{  \vec{\nabla}^2 A }{A} + V +\frac{\partial S}{\partial t} &=&0\, , \label{hjb1} \\
\frac{1}{m}\vec{\nabla} \cdot \left(A^2 {\vec{\nabla}S}\right)  +\frac{\partial A^2}{\partial t}&=&0\, . \label{hjb2}    
\end{eqnarray}

The first equation \eqref{hjb1} is sometimes called the quantum Hamilton--Jacobi equation for the external potential $V(\vec{r})$, which is a modified version of its classical counterpart \cite{holland,wyatt}. The second equation \eqref{hjb2} is the probability conservation (or continuity) equation. The classical Hamilton--Jacobi equation appears modified by the addition of the Bohm potential $V^{Bohm}(\vec{r},t)$ defined by
\begin{equation}
V^{Bohm} (\vec{r},t) \equiv  - \frac{\hbar^2}{2m}\frac{  \vec{\nabla}^2 A }{A}\ . \label{VB2}    
\end{equation}
The Bohm potential does not vanish, in general. There is a family of external potentials \cite{ha20} which admit some solutions for which the Bohm potential vanishes, but even for those potentials, most of their solutions produce non--vanishing Bohm potentials. The vast majority of the potentials have only non--vanishing Bohm potential solutions.\\

The Schrödinger equations for the free particle admit solutions with vanishing Bohm potentials such as constant amplitude plane waves, but the free particle Schrödinger equation also admits solutions with non--vanishing Bohm potentials such as \eqref{ai1} found by Berry and Balasz in 1979 \cite{berry}, written in terms of the Airy function denoted by $Ai$, 

\begin{eqnarray}\label{ai1}
\Psi_{Airy}(x,t)&=&{Ai} \left(\frac{B}{\hbar^{2/3}} \left(x-\frac{B^3}{4m^2} t^2 \right)\right) \nonumber\\
&\times &e^{\frac {i B^3 t\left (6 m^2 x-B^3 t^2 \right)} {12 m^3 \hbar}}\ .
\end{eqnarray}

The above solution describes the propagation of {\it {accelerating free particles}}.\\

In fact, classically, the external potential $V(\vec{r},t)$ determines completely the acceleration of a particle of mass $m$, while in quantum mechanics, the Bohm potential $V^{Bohm}(\vec{r},t)$, which depends on the functional form of the wave function solution, modifies the classical acceleration. The ``quantum'' acceleration $\vec{a_Q}(\vec{r},t)$ may be computed by using (the product of minus one divided by the mass $m$ times)  the gradient of the ``quantum'' potential $V_Q(\vec{r},t)$ which is obtained by adding the Bohm potential $V^{Bohm}(\vec{r},t)$ to the classical external potential $V(\vec{r})$    
 
\begin{eqnarray}\label{acc}
\vec{a_Q}(\vec{r},t)&=&-\frac{1}{m}  \vec{\nabla}V_{Q}(\vec{r},t) \nonumber \\
&=&-\frac{1}{m}  \vec{\nabla}(V(\vec{r}) + V^{Bohm}(\vec{r},t)) \nonumber \\
&\equiv& \vec{a_c}(\vec{r})+\vec{a}^{Bohm}(\vec{r},t) \ .
\end{eqnarray}
These results agree with the expression found for the velocity computed as $\vec{p}/m= \vec{\nabla} S/m$ as recently done in \cite{ha20}. Moreover, they have been experimentally confirmed by using light in optics in 2007 \cite{sivi} and by using electrons in quantum mechanics in 2013 \cite{bloch}.\\

The quantum behavior is therefore determined by both the external potential $V(\vec{r})$ which is unique for a given problem {\it {and}} the Bohm potential which depends on the wave function under consideration \cite{hams21}.\\

\section{Justification of Bohr's Hypothesis}

Take the Schrödinger equations for the hydrogen atom
\begin{eqnarray}\label{sch1}
\left(-\frac{\hbar^2}{2m} \vec{\nabla}^2 -\frac{e^2}{4 \pi \epsilon_0 r} - i \hbar \frac{\partial}{\partial t} \right)  \Psi(\vec{r},t)&=&0\ ,\ \ \ \  {\text {and}}\nonumber \\
\left(-\frac{\hbar^2}{2m} \vec{\nabla}^2 -\frac{e^2}{4 \pi \epsilon_0 r} + i \hbar \frac{\partial}{\partial t} \right)  {\Psi}^{*}(\vec{r},t)&=&0\ . \end{eqnarray}
\noindent which, for time--independent potentials, is usually solved by separation of variables writing

\begin{equation}
\Psi(\vec{r},t)= \psi(\vec{r}) e^{\frac{-iEt}{\hbar}}  
\end{equation}

\noindent where $\psi(\vec{r})$  satisfies

\begin{equation}\label{Schnot}
-\frac{\hbar^2}{2m} \vec{\nabla}^2  \psi_n(\vec{r}) -\frac{e^2}{4 \pi \epsilon_0 r} \psi_n(\vec{r}) - E_n  \psi_n(\vec{r}) =0\ , 
\end{equation}

\noindent and consider the solutions for the Hamiltonian eigen problem, $E_n$ and $\psi_n(\vec{r})$, by using spherical coordinates $r$, $\theta$ and $\phi$,

\begin{eqnarray}\label{ha}
E_n&=&-\frac{m e^4}{32 \pi^2 \epsilon_0 \hbar^2}\frac{1}{n^2},\nonumber \\
\psi_{nlm}(r,\theta,\phi)&=&\sqrt{\Big(\frac{2}{na}\Big)^3 \frac{(n-l-1)!}{2n(n+l)!}}\nonumber \\
&\times&e^{-\rho/2} \rho^l L^{2l+1}_{n-l-1}(\rho)Y^m_l(\theta,\phi)\ ,
\end{eqnarray}
\noindent where 
\begin{equation}\label{rho}
\rho=\frac{2r}{na}    
\end{equation}
\noindent and $a$ is the Bohr radius
\begin{equation}\label{a0}
 a=\frac{4\pi \epsilon_0 \hbar^2}{me^2}\ .   
\end{equation}

The functions $L^{2l+1}_{n-l-1}(\rho)$ are the generalized Laguerre polynomials of degree $n-l-1$ and $Y^m_l(\theta,\phi)$ are the spherical harmonics for angular momentum $l$ and magnetic number $m$.\\

Compute the Bohm potential ${V^{Bohm}}_{ nlm}(r,\theta,\phi)$ for $\psi_{nlm}(r,\theta,\phi)$
\begin{equation}
{V^{Bohm}}_{ nlm}(r,\theta,\psi) =-\frac{\hbar^2}{2m}\frac{\vec{\nabla}^2 \psi_{nlm}(r,\theta,\phi) }{\psi_{nlm}(r,\theta,\phi)} \ .  
\end{equation}
Note that there is no harm in dividing by $\psi_{nlm}(r,\theta,\phi)$ because Eq.\eqref{Schnot} guarantees that the numerator in the Bohm potential definition has, at least, the same nodes that $\psi_{nlm}(r,\theta,\phi)$ has.\\

One gets
\begin{eqnarray}
{V^{Bohm}}_{ nlm}(r,\theta,\phi)&=& \frac{e^2}{4 \pi \epsilon_0 r}-\frac{m e^4}{32 \pi^2 \epsilon_0 \hbar^2}\frac{1}{n^2}\ ,\nonumber \\
&=& -V(\vec{r})+E_n \ .
\end{eqnarray}
Therefore, the quantum potential $V_{Qnlm}(r,\theta,\phi)$ is
\begin{equation}
 V_{Qnlm}(r,\theta,\phi)= V(\vec{r})+ {V^{Bohm}}_{nlm}(r,\theta,\phi) =E_n \ ,
\end{equation}
\noindent and, consequently, {\it{the quantum acceleration $\vec{a}_Q$ vanishes for all of the eigenstates of the Hamiltonian of the hydrogen atom}}. Hence, the quantum acceleration of the electrons in their allowed ``orbits'' vanishes and thus, the electrons on those ``orbits'' do not radiate, which justify, {\it{a posteriori}}, Bohr's assumption.\\

As a matter of fact, the claim in italics in the preceding paragraph is absolutely general in the following sense.\\

Consider the (time--independent) eigenfunctions $ \psi_n(\vec{r})$ of a quantum mechanical Hamiltonian for any classical potential $V(\vec{r})$ with eigenvalues $E_n$, where $n$ denotes collectively all of the quantum numbers (note that the eigenfunctions $\psi_n(\vec{r})$ play the role of the amplitude $A(\vec{r})$ in the description of the Madelung--Bohm approach, understanding that for the eigenfunctions of the Hamiltonian, the full time--dependence of the wave function appears in the phase $S(t)$ only,

\begin{equation}\label{Sch1}
-\frac{\hbar^2}{2m} \vec{\nabla}^2  \psi_n(\vec{r})+ V(\vec{r}) \psi_n(\vec{r}) - E_n  \psi_n(\vec{r}) =0\ , 
\end{equation}
\noindent divide by $\psi_n(\vec{r})$ and compute the quantum potential $ {V_{Q}}_{n}(\vec{r})={V^{Bohm}}_{n}(\vec{r})+ V(\vec{r})$ to get
\begin{equation}\label{Sch2}
{V_{Q}}_{n}(\vec{r})= E_n \ . 
\end{equation}
for all the eigenfunctions of the Hamiltonian for any classical $V(\vec{r})$. There is no need to know the explicit expressions for the wave functions. This proof ends the quantum mechanical justification of Bohr's hypothesis.\\

It is important to recall that for these eigenfunctions, the probability current $\vec{j}$
\begin{equation}\label{j}
\vec{j}=\frac{\hbar}{2mi}\big({\Psi}^{*}(\vec{r},t)\vec{\nabla}{\Psi(\vec{r},t})-{\Psi(\vec{r},t})\vec{\nabla}{\Psi}^{*}(\vec{r},t)\big)
\end{equation}
\noindent as well as the velocity $\vec{v}=\vec{\nabla}S(t)/m$ are equal to zero which, of course, agrees with the fact that the quantum acceleration ${\vec{a}}_Q$ vanishes.\\

To try to understand the agreement between ``quantum'' acceleration and velocity better, it is perhaps worthwhile noting that taking the gradient of Eq. \eqref{hjb1} with the usual Hamilton-Jacobi identifications $\vec{\nabla} S= \vec {p}$ and $\vec {v}=\vec {p}/m$ (where $\vec{p}$ is the momentum) yields

\begin{eqnarray}\label{newton1} 
\vec{v} \cdot \vec{\nabla} \vec {p}+ \frac{\partial \vec {p}}{\partial t} + \vec{\nabla}\big(-\frac{\hbar^2}{2m}\frac{  \vec{\nabla}^2 A }{A} + V \big)  &=&0\ ,   
\end{eqnarray}

\noindent which, using the definition of $\frac{D}{Dt}$ as the usual convective or material derivative \cite{mad},

\begin{equation}\label{conv}
\frac{D \vec {p}}{Dt}\equiv\vec{v} \cdot \vec{\nabla} \vec {p}+\frac{\partial \vec {p}}{\partial t}     
\end{equation}
\noindent may be rewritten as

\begin{eqnarray}\label{newton2}
\frac{D \vec {p}}{Dt} + \vec{\nabla} V_Q  &=&0\ .    
\end{eqnarray}

Note that the previous result is also valid in classical mechanics, the only difference being the addition of the Bohm potential to the external potential to define $V_Q$ in the quantum mechanical setting.\\

Equation \eqref{newton1} is formally equivalent to Euler equation for fluids where the Bohm potential plays the r\^ole of the pressure. This equation, in conjunction with the continuity equation \eqref{hjb2} complete the ``fluid--like'' description of quantum mechanics.\\

\section{Justification of the formula for the radii of the circular orbits}

The very well known Bohr model relation for the radii of the $n^{th}$ allowed radius circular orbit $a_n$ is

\begin{equation}\label{radii}
a_n = n^2 a\ ,    
\end{equation}

where $a_1=a$ is the Bohr radius given in \eqref{a0}.\\

Consider now the ${\Psi}_{nlm}(r,\theta,\phi,t)$ wave function written in terms of a product of functions
\begin{equation}
{\Psi}_{nlm}(r,\theta,\phi,t)= {R}_{nl}(r)Y^m_l(\theta,\phi)e^{\frac{-i E_n t}{\hbar}}
\end{equation}
\noindent where
\begin{equation}\label{R}
{R}_{nl}(r)= \sqrt{\Big(\frac{2}{na}\Big)^3 \frac{(n-l-1)!}{2n(n+l)!}}
\ e^{-\rho/2} \rho^l L^{2l+1}_{n-l-1}(\rho)   
\end{equation}
\noindent with $\rho$ defined in \eqref{rho}.\\

Define the radial distribution function $P_{nl}(r)$, which is proportional to the probability that the electron may be found in a thin spherical shell of radius $r$ by
\begin{equation}\label{P}
{P}_{nl}(r)= r^2 {{R^2}}_{nl}(r)\ .
\end{equation}
The maximum of ${P}_{nl}(r)$ is given by the condition
\begin{equation}\label{Pmax}
\frac{d {P}_{nl}(r)}{dr}= 0\ .
\end{equation}

It is not difficult to show that the maximum value of ${P}_{n\ n-1}(r)$, for $l=n-1$, is attained at the point 
\begin{equation}\label{Pmax}
r_{max}= n^2 a\ ,
\end{equation}

\noindent which is exactly the result obtained using Bohr's model.\\ 

\section{Summary}

The results stated in the preceding sections confirm the basic assumption used to construct Bohr's atomic model and recover its main outcomes, using standard quantum mechanics theory and therefore, provide an {\it{a posteriori}} justification of Bohr's model.\\

\end{document}